\documentclass[graybox,natbib,nosecnum]{svmult}
\bibpunct{(}{)}{;}{a}{}{,} 

\pdfoutput=1   

\usepackage{mathptmx}       
\usepackage{helvet}         
\usepackage{courier}        
\usepackage{type1cm}        

\usepackage{makeidx}         
\usepackage{graphicx}        
\usepackage{multicol}        
\usepackage[bottom]{footmisc}
\usepackage[normalem]{ulem}	
\usepackage{hyperref}  

\usepackage{soul}   

\makeindex             

\begin{document}

\title*{Exoplanet Atmosphere Measurements from Direct Imaging}
\author{Beth A. Biller and Micka\"el Bonnefoy}
\institute{Beth A. Biller \at Institute of Astronomy, University of Edinburgh, Edinburgh, UK
 \email{bb@roe.ac.uk}
\and Micka\"el Bonnefoy \at University of Grenoble \email{mickael.bonnefoy@univ-grenoble-alpes.fr}}
%
%
\maketitle

\abstract{In the last decade, about a dozen giant exoplanets have been directly imaged in the IR as companions to young stars.  With photometry and spectroscopy of these planets in hand from new extreme coronagraphic instruments such as SPHERE at VLT and GPI at Gemini, we are beginning to characterize and classify the atmospheres of these objects.  Initially, it was assumed that young planets would be similar to field brown dwarfs, more massive objects that nonetheless share similar effective temperatures and compositions.  Surprisingly, young planets appear considerably redder than field brown dwarfs, likely a result of their low surface gravities and indicating much different atmospheric structures.  Preliminarily, young free-floating planets appear to be as or more variable than field brown dwarfs, due to rotational modulation of inhomogeneous surface features.  Eventually, such inhomogeneity will allow the top of atmosphere structure of these objects to be mapped via Doppler imaging on extremely large telescopes.  Direct imaging spectroscopy of giant exoplanets now is a prelude for the study of habitable zone planets. Eventual direct imaging spectroscopy of a large sample of habitable zone planets with future telescopes such as LUVOIR will be necessary to identify multiple biosignatures and establish habitability for Earth-mass exoplanets in the habitable zones of nearby stars.}

\section{Introduction }

Since 1995, more than 3000 exoplanets have been discovered, mostly via indirect means, ushering in a completely new field of astronomy.  In the last decade, about a dozen planets have been directly imaged, including archetypical systems such as HR 8799bcde and $\beta$ Pic b \citep{Mar08, Maro10, Lag10}.  Progress in discovering exoplanets has been dramatic, even yielding planets with masses similar to or somewhat larger than the Earth in the “habitable zone” of their parent star \citep[c.f.~][]{Jen15,Ang16,Kane16}, i.e. at distances from their parent star that may allow liquid water on their surfaces.  However, whether or not these planets are capable of hosting life remains an open question -- the habitable zone planets known to date have only been detected indirectly, meaning we cannot probe their atmospheric composition.  Atmospheric characterization of exoplanets via spectroscopy and photometry is necessary to determine if so-called habitable zone planets are truly habitable.  Only through studying the atmospheres of exoplanets across a wide parameter space can we truly understand the underlying physical properties of these atmospheres.  There are only two methods -- direct imaging and transit spectroscopy -- that yield photons from the planet itself, enabling direct measurement of atmospheric properties.  In this review, we will discuss the current state of the art regarding direct imaging characterization of giant exoplanet atmospheres as well as prospects for future studies of lower mass and even habitable exoplanets with direct imaging.

Such spectroscopic techniques are already in use today to characterize the atmospheres of highly irradiated transiting extrasolar giant planets, mini-Neptunes, and super-Earths \citep{Sing16,Krei14}, as well as wide, young giant planets \citep{Mac15,Bon16,Ch17}.
Exoplanet spectroscopy has revealed that clouds (formed of silicate condensates in the case of wide giant exoplanets) are likely common both for transiting and directly imaged planets \citep{Krei14,Ske12,Ske14}.  While transit spectroscopy is a highly valuable technique for studying exoplanet atmospheres, due to viewing geometry, only $\sim$10$\%$ of planets will have observable transits, and even less for habitable zone planets, which are on average further from their parent star. Thus direct imaging will provide a more complete census of the atmospheres of planets, particularly those around higher mass stars where transits of habitable zone planets are too infrequent to characterize systems on reasonable timescales. Direct imaging spectroscopy of planets also literally provides a different angle compared to the currently dominant transmission spectrum method of studying transiting exoplanet atmospheres. Transmission spectra probe the upper layers of atmospheres at glancing incidence, whereas direct imaging probes straight into these atmospheres, reaching deeper and enabling higher S/N detection of interesting spectral features such as biosignatures \citep{Morl15}.  

The current cohort of directly imaged planets generally have estimated masses $>$2 M$_{Jup}$ and orbit young stars ($<$300 Myr).  These planets have been imaged primarily in the near- (1-3 $\mu$m) and mid-IR (3-5 $\mu$m), near the blackbody peak of their own thermal emission.  At these young ages, the planets are still quite warm (effective temperatures, henceforth $T_\mathrm{eff}$, between 400 and 2000 K) and are hence significantly more self-luminous than at field ages.  For such young planets observed at infrared wavelengths, a star-planet contrast ratio in the IR of $\sim10^4 -10^7$ is expected \citep{Beu08,Mac14}.  Around nearby, young stars (distances $<$100 pc), these planets are found at projected separations of 0.05-1", although very wide exoplanet companions have also been imaged, up to separations of $\sim$2000 AU \citep{Nau14}.  Such contrasts and resolutions are well within the reach of the current generation of extreme adaptive optics coronagraphs such as SPHERE at VLT \citep{Beu08} and GPI at Gemini \citep{Mac14}.

Indeed, the first few directly imaged exoplanets -- specifically, HR 8799bcde, $\beta$ Pic b, HD 95086b, and GJ 504b \citep{Mar08, Maro10, Lag10, Ram13, Kuz13} -- were actually discovered with facility AO systems without coronagraphs.  The advent of extreme adaptive optics coronagraphs with integral field spectrographs has enabled the characterization of spectra and atmospheres for these early discoveries as well as the discovery of additional exoplanet companions \citep{Mac15,Ch17}.  The initial results from this characterization have been surprising -- young exoplanet companions possess significantly redder colors than old field brown dwarfs with similar effective temperatures, suggesting that the much lower surface gravity of the young objects plays a role in atmospheric properties.  Additionally, a cohort of young, free-floating planetary mass objects with ages, effective temperatures, estimated masses, and properties similar to the cohort of companion objects has been identified \citep{Gag14}.  This allows us to characterize a much larger sample of objects with comparable properties.

In the next decades, we will characterize this cohort of massive planets in detail, especially at previously inaccessible wavelengths, thanks to the advent of JWST.  Space missions such as WFIRST-AFTA will enable the direct detection of Neptune and lower mass objects around Sun-like stars.   Extremely large telescopes may image habitable zone exoplanets around nearby M stars.  Finally, in the longer term future, LUVOIR may be able to both detect and characterize Earth-like habitable zone exoplanets around solar-type stars.  

In this article, we will summarize the current state of the art in terms of our understanding of young giant planet atmospheres via direct imaging.  We will also discuss prospects for deeper characterization of known exoplanets as well as direct imaging of lower mass, closer-in exoplanets in the short term with JWST and in the longer term with the extremely large telescopes, WFIRST-AFTA, HabEx, and LUVOIR.  Finally, we will discuss prospects for detecting and characterizing habitable zone, Earthlike planets both from the ground and space. 

\section{Methods}

Direct imaging of exoplanets is technically very challenging -- imaging even the brightest, young exoplanet companions requires contrasts $>$10$^5$ and resolution into the inner arcsec around the star.  Such resolution is possible either from space (where one will be by definition diffraction-limited) or with adaptive optics from the ground.  However, even in the diffraction-limited case or after adaptive optics correction, significant quasi-static speckles remain in the inner arcsec, producing a limiting noise floor above the contrasts necessary to image exoplanets \citep{Rac99}.  Thus, various speckle mitigation techniques have been developed to overcome this speckle noise floor.  

Numerous speckle-suppression techniques have been proposed and implemented; what these techniques share in common is that they use instrumental methods to distinguish speckles from faint companions.  The technique that has become the standard for the field is known as azimuthal differential imaging (henceforth ADI) and uses sky rotation to decorrelate real objects from speckles \citep{Mar06}.  Instead of using the telescope rotator to place north up and east left in all images, in an ADI sequence, the rotator is turned off and the field is allowed to rotate with parallactic angle.  The speckles, which are instrumental, will continue to appear in roughly the same positions in the image; a real companion or other object in the field, however, will be modulated by the change in parallactic angle.  Multiple algorithms have been developed \citep[c.f.~LOCI,~KLIP,~PCA,~Andromeda,~][]{Laf07, Sou12, Ama12, Can15} to build a model of the speckle pattern in the image on a frame by frame basis, subtract it from each frame, and then stack to produce a "de-speckled" image.    

If the adaptive optics correction is high enough (e.g. Strehl ratio $>$0.9, where Strehl ratio=1 would be a perfect diffraction limited image), a coronagraph will also provide significant gains in achievable contrast \citep{Kuc02}.  Today, the standard in the field is coronagraphic imaging in tandem with ADI observations.  To date, most of the work in this field has been ground-based, as that is where coronagraphs with tight inner working angles are available -- HST, while diffraction limited, does not have appropriate coronagraphs for imaging close-in planets.  In addition, various ground-based spectroscopic options have become available in the last few years.  SPHERE at VLT, GPI at Gemini and now SCExAO+Charis at Subaru \citep{Grof16} all combine high-contrast coronagraphs with integral field spectrographs to enable low resolution spectroscopy ($R\sim$30-83) of exoplanet companions.  Integral field spectrographs use image slicers to obtain simultaneous high contrast images in multiple wavelengths, in other words, low resolution spectra.  For brighter companions, long-slit spectroscopy can enable higher resolution spectroscopy in combination with high-contrast imaging -- e.g. the long-slit spectroscopy modes available with SPHERE at VLT \citep{Vig08, Mair15}.  In a few cases, it is possible to acquire extremely high contrast spectroscopy in tandem with adaptive optics imaging -- this has enabled the first measurement of a rotation period for an extrasolar planet \citep{Sne14}.  

\section{Current results}

\begin{figure}
\includegraphics[scale=0.8]{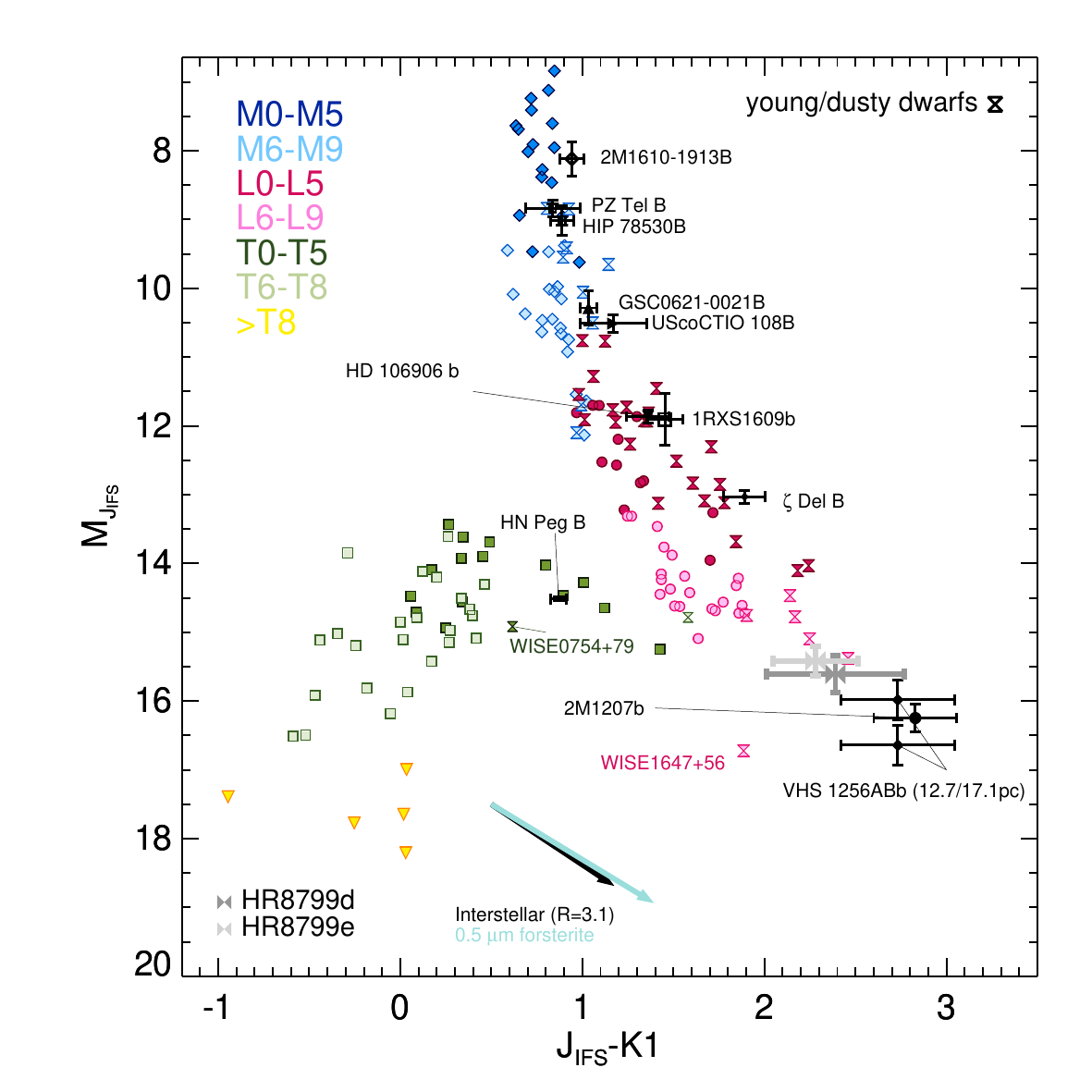}
\caption{Color-color diagrams for HR 8799bcde, free-floating planetary mass objects and wide planetary mass companions. The $J_{IFS}$ magnitudes correspond to a virtual filter encompasing the 1.20-1.32 $\mu$m range of the integral field spectrograph (IFS) of SPHERE. The K1 filter has a 2.06-2.16 $\mu$m passband. Field dwarfs are overplotted for comparison. All exoplanets and brown dwarfs begin life as  M-type objects, cooling and moving red-wards to the L spectral type, then transitioning to the comparatively blue T spectral type (due to strong methane absorption that removes flux predominantly in the K band).  Compared to field brown dwarfs, young planetary mass objects retain red colors and L spectral types down to much lower $T_\mathrm{eff}$.
}
\label{fig:1}       
\end{figure}

Since direct imaging yields luminosities of planets but does not immediately provide a measurement of mass (as the combination of radial velocity and transit detection of the same planet does), objects are studied and classified according to their measured effective temperature $T_\mathrm{eff}$. Old field brown dwarfs, young free-floating planetary mass objects, and exoplanet companions to stars are described using the same set of spectral types.  Unlike stars which remain at the same effective temperature for millions or billions of years due to nuclear core fusion, brown dwarfs and planetary mass objects alike cool monotonically with age, beginning life as hot
M-type objects, cooling to the L spectral type (characterized by very red near-IR colors and silicate condensate clouds \citep{Kir05}, the T spectral type  \citep[characterized by blue near-IR colors and strong methane absorption at 1.6 $\mu$m and 2.2 $\mu$m,][]{Kir05} and eventually to the  very cool Y spectral type \citep[][see Fig.~\ref{fig:1}]{Cus11, Kir12}.
Thus, there is an age / mass / temperature degeneracy for these
objects.   Additionally, it has become clear that low surface gravity in younger and lower mass objects can significantly affect the spectra of these objects and also the
T$_\mathrm{eff}$ at which these objects
transition between spectral types \citep{Bar11}.

The number of directly imaged exoplanets remains small, both because of the technical difficulty of their detection, and also because initial surveys have demonstrated that giant exoplanets are rare at separations$>$30 AU, where we can readily detect them \cite[c.f.~][]{Bil13b,Bra14}.  However, a larger, statistically significant number of objects with similar properties are available to study if we consider as well free-floating planetary mass objects with similar ages.  Many direct-imaging planet host stars known to date (e.g. HR 8799, $\beta$ Pic) are members of nearby young moving groups -- aggregations of young stars (10-300 Myr) in the solar neighborhood that were formed in the same star-forming cloud and still share similar sky motions, even though the cloud materials have long since dissipated \citep[][~also,~for~an~example~of~a~recent~membership~search~of~such~a~group]{Zuck04,Sch12}.
In the last few years, searches for new moving group members using advanced Bayesian techniques have begun to identify planetary mass or very low mass brown dwarf ($<25 M_{Jup}$) members of these groups \cite[see~c.f.~][]{Gag14,Gagn15,Fahe16}.  On order 30 such objects have been identified to date.  These objects do not necessarily share the same formation mechanism as young exoplanet companions to stars, but they do share similar ages, masses, and compositions and are unobscured by a nearby bright star.  Thus, they serve as important and much more easily studied proxies to directly imaged planets.  Similarly to directly imaged planets, young free-floating planetary mass objects possess significantly redder colors than field brown dwarfs with similar spectral types.   For the rest of this section, we consider results from free-floating planetary mass objects alongside results from bonafide planetary mass companions.

\subsection{Initial photometric measurements}

Given the age / temperature / luminosity degeneracy for substellar objects, it was initially expected that directly imaged exoplanets should have similar colors and spectral types as T spectral type brown dwarfs, as both directly imaged exoplanets and more massive but older T type brown dwarfs should have expected $T_\mathrm{eff} < 1400 K$ and similar compositions.  T-type brown dwarfs possess notable methane absorption features at 1.6 and 2.2 $\mu$m; early searches for 
exoplanet companions specifically targeted deep methane absorption expected in these objects \citep{Bil07}.  However, in contrast to expectation, the first planetary mass companions imaged had extremely red colors relative to field dwarfs with similar $T_\mathrm{eff}$ and no methane absorption.  In Fig.~\ref{fig:1}, we plot $J-K$ color-magnitude diagram and report the photometry of young companions, including the those of the planets HR8799 d and e, as well as a subset of other young free-floating and companion planetary mass objects.  M, L, and T field brown dwarfs are plotted as well.  The young planetary mass objects plotted clearly have considerably redder $J-H2$ colors than their field counterparts with similar $T_\mathrm{eff}$.  These objects, however, also must have lower surface gravities compared to the field dwarfs, which should affect atmospheric structure.  Various groups have suggested that thick and/or patchy clouds are necessary to explain the extremely red photometry measured for these objects \citep{Bar11, Mad11}.

\subsection{High contrast spectroscopy}

With the advent of high-contrast imagers with integral field spectrographs such as SPHERE and GPI, low resolution (R$\sim$30-83) near-IR spectra have been acquired for several bright planets \citep{Bon16,Zur16,Bon14,Ing14,Chi15,DeR16, Mac15, Sam17}, allowing for a more detailed inspection of atmospheric properties.  We summarize these results, starting with the highest effective temperature planets and continuing on with decreasing temperature. 

$\beta$ Pic b is a $\sim$8-13 $M_{Jup}$ companion 8-10 AU from the $\sim$20 Myr intermediate mass star $\beta$ Pic.  Given the higher mass of its primary relative to the Sun, it is very much like a higher mass analogue to Saturn in our own solar system and could have formed in a similar way as Jupiter. In 2014, the Gemini Planet Imager instrument (GPI) was used to capture the first low-resolution near-infrared ($1.1-1.8 \mu$m) spectra of the exoplanet \citep{Bon14, Chi15}. The spectrum displays water band absorptions and is characteristic of young low-gravity objects. This spectrum and additional photometry from 2 to 5 $\mu$m shows that the planet has an early L spectral type, an effective temperature around 1700 K \citep{Bon14,Morz15}, and a radius of $\sim 1.5 R_{Jup}$.

\begin{figure}
\begin{centering}
\includegraphics[scale=0.63]{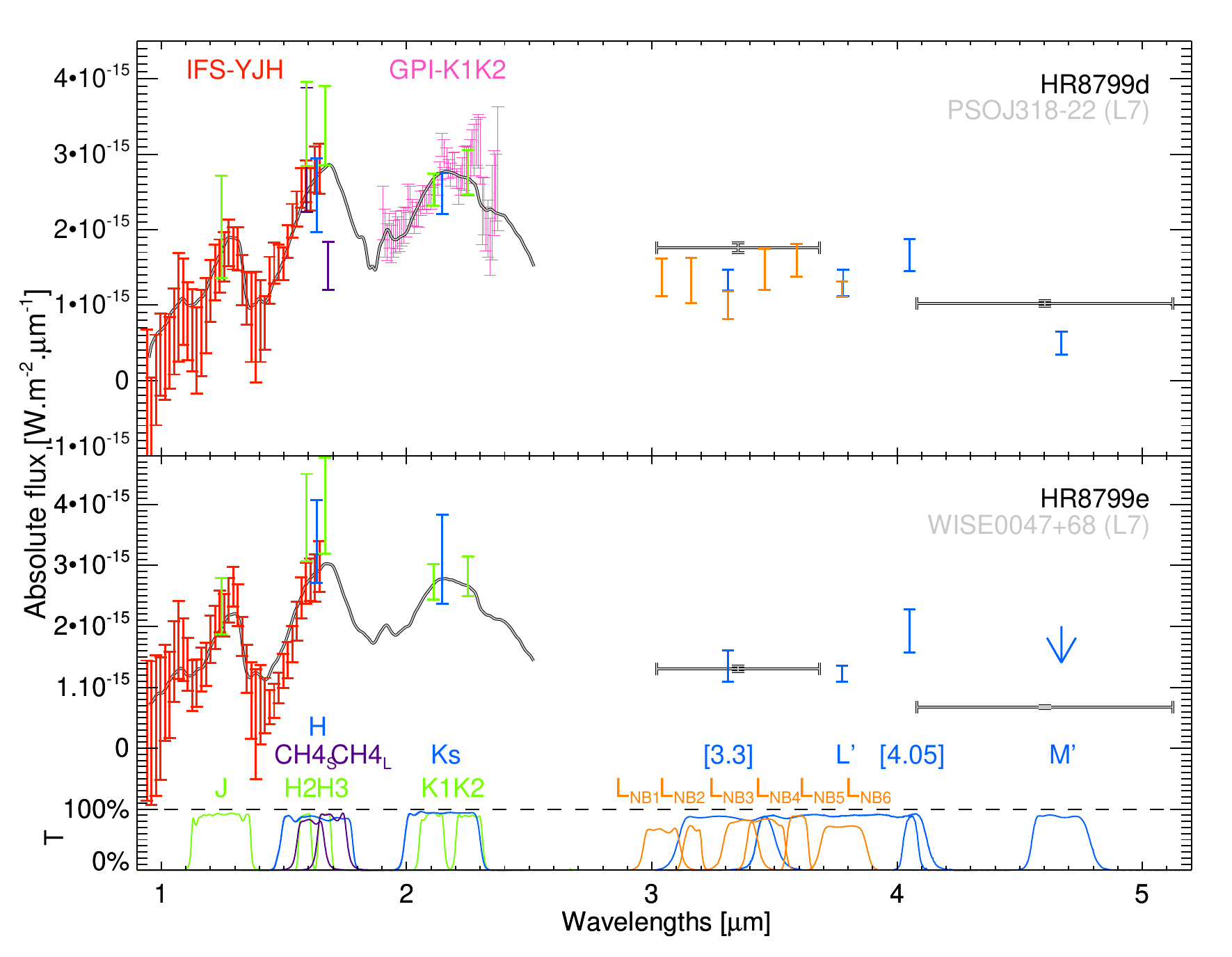}
\end{centering}
\caption{Near-IR spectra and mid-IR spectrophotometry for the planets HR 8799d and e. with data from \cite{Oppo13,Ske12,Ske14,Ing14, Bon16, Zur16}.}
\label{fig:HR8799}       
\end{figure}

The four planets orbiting the intermediate mass star HR 8799 all have L/T spectral types and estimated masses between 5  and 7 $M_{Jup}$.  They offer the  opportunity to perform comparative planetology beyond our Solar System.  In the past few years, images of that system have been gathered from 0.9 to 5 $\mu$m. Using spectrophotometry at mid-IR wavelengths from 3-5 $\mu$m, \cite{Ske12} and \cite{Ske14} find that patchy cloud models and/or non-equilibrium chemistry are necessary to model the spectral structure at these wavelengths for HR 8799b, c, and d. More recently, the instruments P1640, SPHERE, and GPI  yielded low-resolution near-infrared (0.9-2.5 $\mu$m) spectra of each planets \citep{Oppo13,Ing14,Zur16} which can be compared to those of known objects (see Fig.~\ref{fig:HR8799}). The spectra of the two innermost planets are similar to those of young and peculiar late-L brown dwarfs identified in the solar neighborhood \citep{Bon16}. This confirms that the peculiar properties of the planet emergent fluxes are related to their low surface gravity. The spectra of the two outermost planets cannot be represented by those of any other known object. But free-floating analogues of HR8799b and c may exist in the Solar neighborhood or in young clusters and remain to be found. 

\begin{figure}
\includegraphics[scale=0.63]{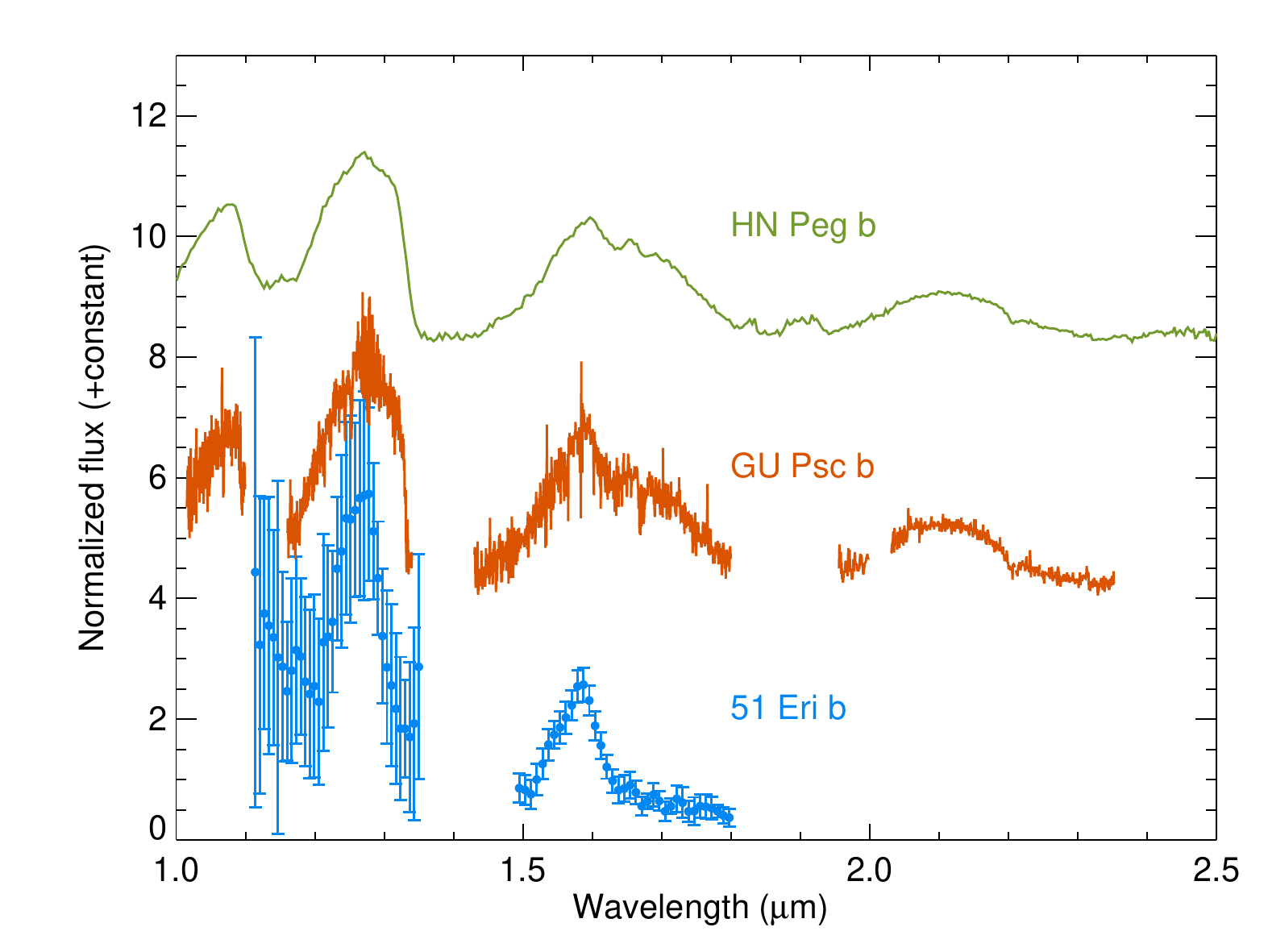}
\caption{Spectra of young companions 51 Eri b, HN Peg b, and GU Psc b taken from \cite{Mac15,Nau14,Luh07} with methane absorption at 1.6 $\mu$m and 2.2 $\mu$m. The young Jupiter analogue 51 Eri b is the first clearly T-type planet.}
\label{fig:Ttypeplanets}       
\end{figure}

The increased sensitivity of the new generation of direct imaging instrument SPHERE and GPI enable detection of fainter and cooler objects closer to their parent stars. 
51 Eri b \citep{Mac15} is the first exoplanet to be identified by these instruments and is also the first planet to clearly exhibit methane absorption features, shown in Fig.~\ref{fig:Ttypeplanets}.  Only two other young wide-orbit companions show similar absorption features \citep[GU Psc b and HN Peg b;][]{Nau14, Luh07}, but is likely that those objects did not form like planets. 51 Eri b is quite a bit cooler than other imaged planets and suggests that the L/T transition occurs at considerably lower $T_\mathrm{eff}$ for low surface gravity planets than for high surface gravity brown dwarfs.

\subsection{Variability}

Most direct imaging observations of exoplanets to date have had short durations of 2 hours or less, and thus probe planet properties in a time-averaged and spatially unresolved manner.
However, new exoplanet imagers such as GPI and SPHERE offer the possibility to directly image exoplanets with cadences of 20-30 minutes, allowing us to search for rotationally modulated variability. Detectable rotationally modulated variability would be produced in giant exoplanets in the presence of: 1) relatively short rotation periods (i.e. observable on the timescales of a
night or two) and 2) surface inhomogeneities, potentially caused by inhomogeneous cloud cover \cite{Apa13} or thermo-chemical instabilities \cite{Tre15,Tre16}.  Both of these
conditions are fulfilled for brown dwarfs and appear also to be fulfilled for young giant planets.

Brown dwarfs are known fast rotators, with periods of 3-20 hours \citep{Zap06}.  Preliminarily, one might expect young giant exoplanets to be somewhat slower rotators, as their younger ages imply that they have not had as much chance to spin up as they contract.  However, initial measurements of rotational period for planetary mass objects (both companions and free-floating) have found periods generally less than 20 hours (in 4 out of 5 cases to date).  The first such measurement, of the giant exoplanet $\beta$ Pic b, derived a rotation period of 7-9 hours for this planet from the Doppler broadening of spectral features in the planet's atmosphere \citep{Sne14}.  From its photometric variability, 2MASSW J1207334-393254b (henceforth 2M1207b), a young planetary mass companion to a 20-30 $M_{Jup}$ brown dwarf, was found to have a rotation period of $\sim$11 hours \citep{Zho16}.  The free-floating planetary mass object PSO J318.5-22 has a rotation period between 5 and 10 hours \citep{Bil15,All16} and the very low mass brown dwarf W0047 ($<20 M_{Jup}$) has a rotation period of $\sim$13 hours \citep{Lew16}.  Of these early measurements, the one outlier is the rather more massive companion GQ Lup B ($\sim40 M_{Jup}$), which is a much slower rotator with a period of 82 days in the case that the orientation of the spin axis is edge-on \citep{Sch16}.  However, taken as a whole, it is not implausible that exoplanet companions other than $\beta$ Pic b may have relatively fast rotation periods.

Surface inhomogeneities provides the second ingredient for quasi-periodic photometric variability.  Such variability is already commonly observed for field brown dwarfs \citep{Rad14a,Wil14,Rad14b,Met15} and is often attributed to patchy thin and thick clouds, especially along the L/T transition \citep{Apa13}, although recently thermochemical instabilities have also been proposed as a source of surface inhomogeneities \citep{Tre15,Tre16}.  Young exoplanets should possess similar clouds  \citep[generally composed of hot silicate particles; ][]{Bar11,Ske12,Ske14}, although the structure of these clouds may vary due to the lower surface gravities of these objects relative to field brown dwarfs.  At least preliminarily, it seems likely that young exoplanets should also be similarly variable.  

Only two known systems have planets sufficiently bright for variability monitoring on reasonable cadences with current instruments, specifically HR 8799bcde and $\beta$ Pic b.  One search for variability in exoplanet companions to main sequence stars has been published to date --   \citet{Apa16} obtained SPHERE science verification data to monitor for variability in HR 8799bcd over several nights in December 2015.  As HR 8799 is only available at reasonable airmasses at the very beginning of the night in December, each observation was only 30 minutes long.  Thus, while \citet{Apa16} demonstrated 
that satellite spot-modulated artificial planet-injection based photometry (SMAP) produced a significant ($\sim$3×) gain in photometric accuracy over standard aperture-based photometry, they did not possess the cadence to detect variability on time scales similar to the expected rotation periods of the planets.   

\begin{figure}
\includegraphics[scale=.4]{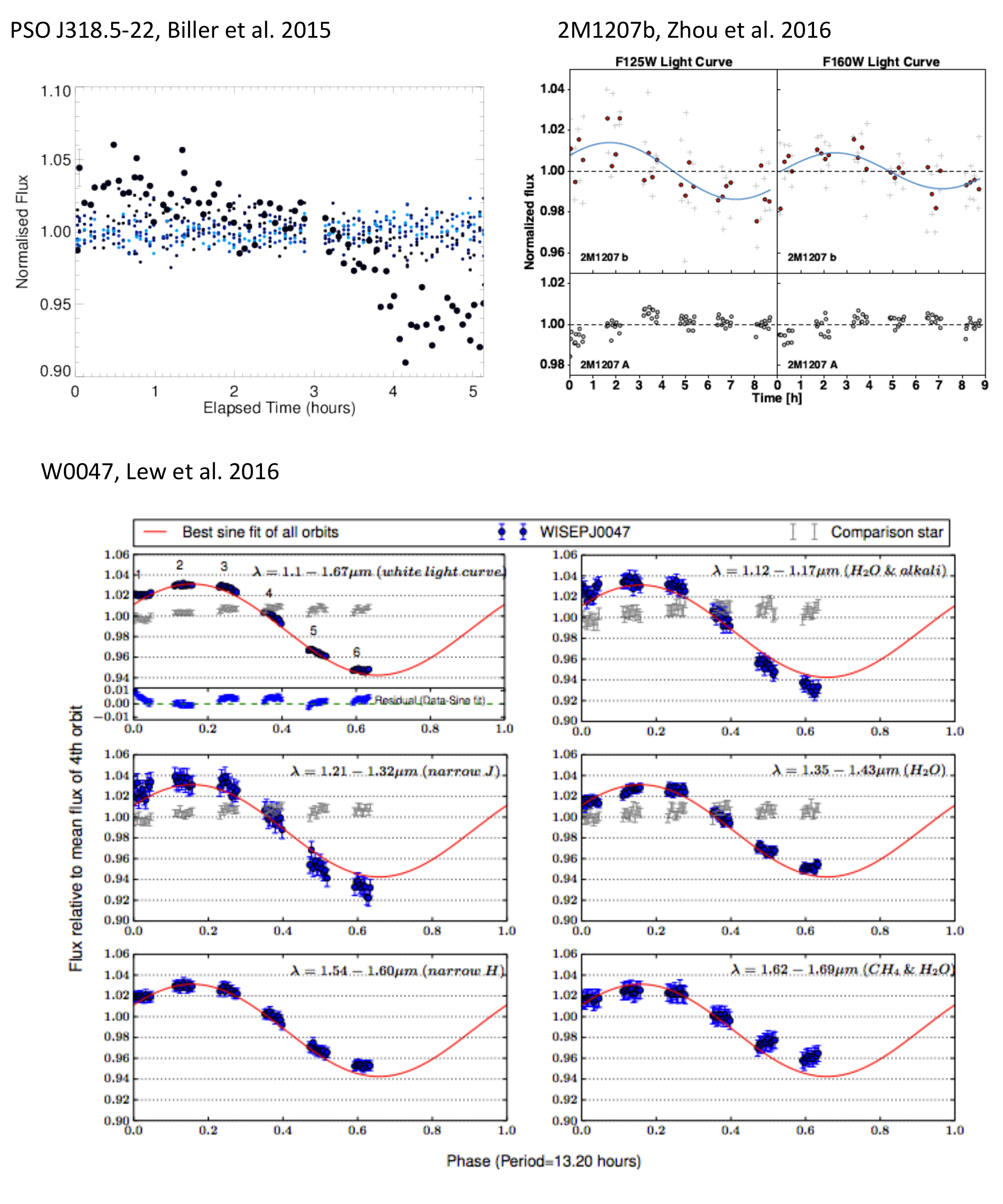}
\caption{Variability of young free-floating planetary mass objects and wide companions, figures from \citet{Bil15,Zho16,Lew16}.  Preliminarily, quasi-periodic variability potentially driven by cloud features appears to be as common or more common in low surface gravity planetary mass objects compared to higher surface gravity field brown dwarfs.}
\label{fig:variability}       
\end{figure}

While the extreme photometric stability expected with JWST will open up a few more systems for variability monitoring, free-floating planetary mass objects and young, very low mass brown dwarfs provide another (considerably larger) sample of similar objects for variability studies. Preliminarily, these objects seem to be as variable or more variable than field brown dwarfs.  \citet{Bil15} published the first detection of such variability for the $\sim 8 M_{Jup}$ free-floating planetary mass object PSO J318.5-22 \cite{Liu13}, with an amplitude of 7-10$\%$ in J.  Soon thereafter, \citet{Lew16} reported high-amplitude variability in the young, very low mass brown dwarf WISEP J004701.06+680352.1 (henceforth W0047, a member of the AB Dor moving group).  Both of these objects have late L spectral types and provide a striking contrast to variability detected to date for field brown dwarfs.  Among field brown dwarfs, L/T transition brown dwarfs display the highest variability amplitudes \citep{Rad14a}, interpreted as break-up of clouds along the L to T spectral type transition producing patchiness and hence variability.  In contrast PSO J318.5-22 and W0047 have late L spectral types, and presumably thick dusty clouds -- but variability amplitudes similar to L/T transition brown dwarfs (and are in fact the highest amplitude variables among the entire L spectral type), suggesting distinctly different atmospheric structure compared to higher-gravity field objects.  Lower level variability with a period of 11 hours has also been detected for 2M1207b, a late-L planetary mass companion to a young brown dwarf \citep{Yan16}.  Lightcurves for variable exoplanet analogues are shown in Fig.~\ref{fig:variability}.

\begin{figure}
\includegraphics[scale=1.5]{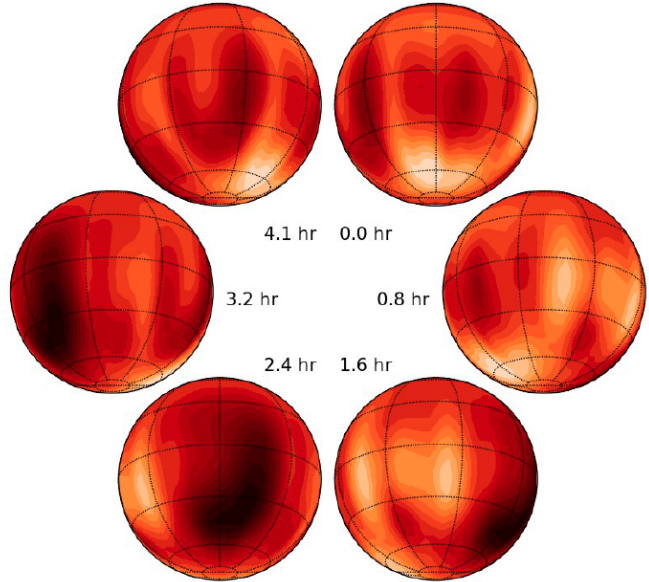}
\caption{Doppler map of the early T variable brown dwarf Luhman 16B, from \cite{Cro14a}.  High-resolution spectrographs on extremely large telescopes will enable similar mapping of directly imaged exoplanet companions.}
\label{fig:Dopplermap}       
\end{figure}

Variability eventually enables mapping of planetary surfaces, as the time-dimension allows us to probe different parts of the surface at different times in the observation.  For brown dwarfs, significant effort has already gone in to reconstructing surface features based on spectroscopic variability monitoring \citep{Apa13, Kar15, Kar16}. Indeed, for the closest brown dwarf to the Earth, the L/T transition binary Luhman 16AB \citep{Luh13}, the surface of the highly variable B component \citep{Gil13,Bil13} has been mapped via the Doppler imaging technique \citep[][]{Cro14a}, presented here in Fig.~\ref{fig:Dopplermap}.
While similar studies for exoplanets are still in their infancy, eventually similar maps will be produced for exoplanets as well -- in particular, the high-resolution spectrograph planned for the METIS instrument on the E-ELT may provide Doppler images of the surface of $\beta$ Pic b, a fast rotator with a 7-9 hour period.

\subsection{Very young accreting planets}

 	Young companions are thought to be surrounded by circumplanetary disks, from which they acquire their mass \citep[e.g.~][]{Quil98}. The accretion of the hydrogen gas contained in these disks produces characteristic emission lines in the companion spectra in the optical ($H_{\alpha}$, 656 nm) and in the near-infrared ($Pa_{\beta}$, 1.282 $\mu$m). Up to now, those signatures have been found for five young (1-11 Myr) companions with masses between 11 and 17 M$_{Jup}$ and large projected separations of several hundreds of au \citep{Seif07,Schm08,Bowl11,Bowl14,Bonn14,2014ApJ...783L..17Z}.
The line intensities were used to estimate gas accretion rate \citep[e.g.][]{2014ApJ...783L..17Z} ranging from  $10^{-11.5}$ to $10^{-9.3} M_{\odot}/year$. Similar accretion rates are found for free-floating brown dwarfs down to 11 $M_{Jup}$ \citep{Natt14,Joer13}.
    
\begin{figure}
\includegraphics[scale=0.5]{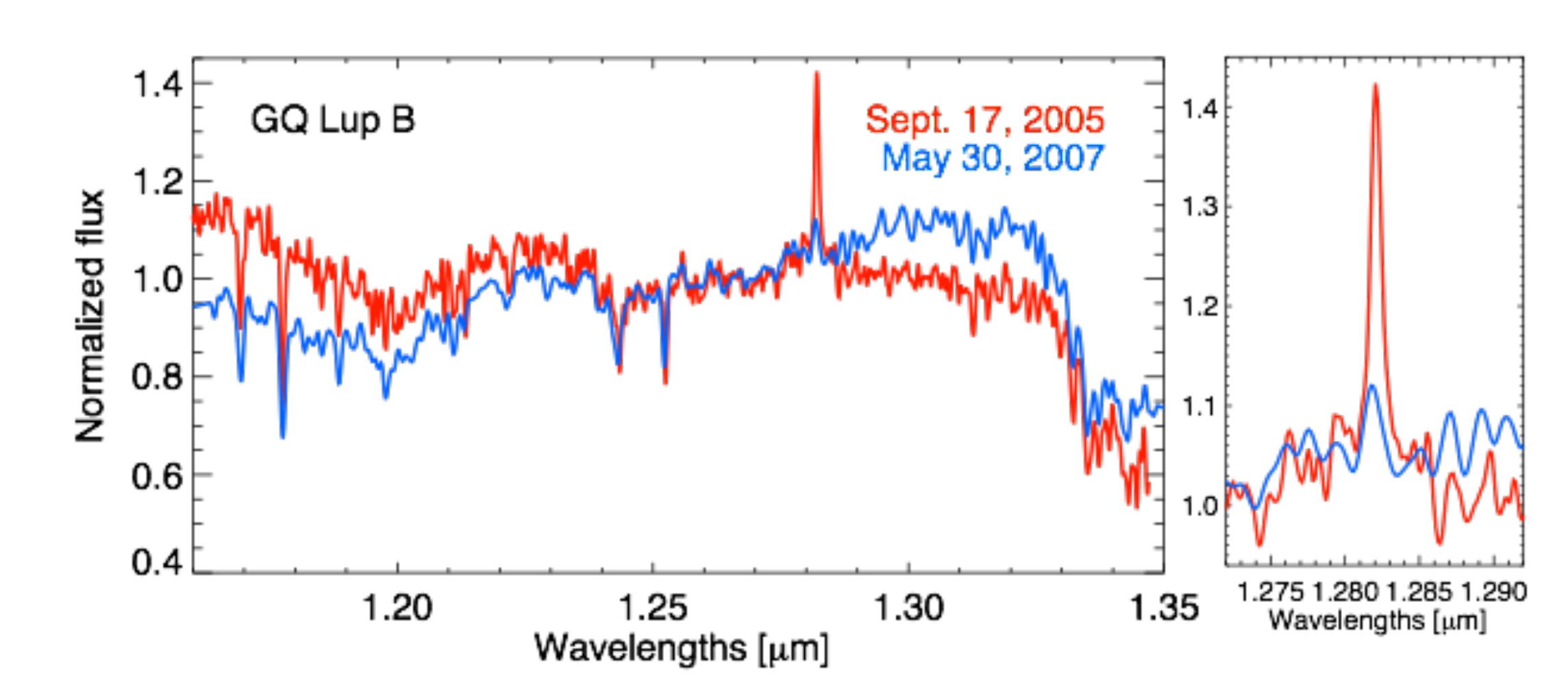}
\caption{Near-infrared spectra of the low-mass companion GQ Lup B. In 2005, the companion spectrum exhibited a strong Paschen $\beta$ emission line at 1.282 $\mu$m \citep{Seif07}. This line had nearly vanished in a spectrum taken in 2007 \citep{Lavi09}.}
\label{fig:evolPaB}       
\end{figure}

    The intensity of those lines can evolve in time (Fig. \ref{fig:evolPaB}). In 2005, the companion GQ Lup b displayed a strong Paschen $\beta$ line which had nearly vanished in 2007 \citep{Seif07,Lavi09}.  The same phenomenon was observed more recently for the companion GSC 06214-00210 b \citep{Lach15} in the near-infrared and in the optical for GQ Lup b \citep{Wu17}. The amplitude and timescales of the line variability could provide precious clues on the way the planets accrete their material, and how their physical properties (temperature, radius, luminosity) evolve in time \citep[up~to~$\sim$100 ~Myr;~][]{Marl07, Bara09}.  
 
   Recent observations with ALMA have set constraints on the mass reservoir surrounding those objects \citep{Cace15,Bowl15,Krau15,Macg17,Wu17}.  The accretion rate and disk mass  are related to the circumplanetary disk lifetime and indirectly determine the formation mode and timescale of the gas giant planets. 
   
   The line emission can be exploited to detect young accreting planets within circumstellar disks. In the optical, the continuum of the planet is faint so that carefully designed narrow-band filters centered on an off the $H_{\alpha}$ line can capture the emission of the planet while removing the stellar flux. This technique has been implemented on the new generation of extreme adaptive optics instruments such as MagAO on the Magellan telescope \citep{Clos13} and SPHERE at the VLT \citep{Beu08}, and has already yielded the detection of an accreting stellar companion to the Herbig star HD142527 \citep{Clos14} and of a Jovian planet around the star LkCa 15 \citep{Sall15}.
        
\section{Prospects for the future}

We have summarized the current state-of-the-art regarding the characterization of exoplanet atmospheres via direct imaging with current (largely ground-based) instruments.  This field is pushed forward by technology development and the advent of new telescopes and instruments.  A few critical developments will produce notable steps forward in this field.  In the next 2 years, the advent of JWST will lead to much more detailed characterization of bright, relatively wide exoplanet companions and free-floating planetary mass objects.  In 10 years, the extremely large telescopes will enable in-depth characterization of Neptune to Jupiter-mass exoplanets and the potential to directly detect super-Earths around very nearby M-stars.  WFIRST-AFTA will directly image super-Earths to Neptune planets and in the very long term future, HabEx and LUVOIR may image the first habitable exoplanets around Sun-like stars.

\subsection{The next 5-10 years}

To date, no directly imaged exoplanet has been characterized at wavelengths longer than 5 $\mu$m. This is because current facilities are ground based and thus suffer from the very high sky background at these long wavelengths (since the blackbody peak for the Earth's atmosphere is at around 10 $\mu$m so literally everything around your telescope is glowing).  To effectively observe at these wavelengths, it is imperative to work from space, above the Earth's atmosphere.  The mid-IR Spitzer space telescope was capable of this during the cooled Spitzer mission, but the coarse resolution of Spitzer and lack of coronagraphs precluded studying close companion objects.  The MIRI instrument in JWST possesses four-quadrant phase masks at 10.65, 11.4 and 15.5 $\mu$m, and a Lyot coronograph at 23 $\mu$m, allowing the study of exoplanet companions at these wavelengths for the first time \citep{Bocc15}.  The four-quadrant phase masks have inner working angles at approximately $\lambda / D$, limiting studies to companions at separations greater than 0.35-0.5".  Nonetheless, this will enable long wavelength photometry of much of the cohort of directly-imaged exoplanet companions.  

The high quality spectra and photometry of giant exoplanets obtained with the JWST will be highly suitable to atmospheric retrieval studies. The retrieval method uses a Bayesian framework (Markov-chain Monte-Carlo, nested sampling) to represent a set of data and associated errors and infer posterior distributions on each free parameter considered in the models. The method has  been successfully applied to the spectra of brown dwarf companions \citep{Line14,Line15,Burn17} and of transiting exoplanets \citep[e.g.~][]{Krei14b,Benn16}. It can set constraints on their atmospheric structures and composition. In the JWST era, the method promises to provide unprecedented clues to the origin of giant planets \citep{Lee13, Lavi16} through the quantification of the abundances of carbon (C), oxygen (O), and hydrogen (H) provided that the formation of the cloud deck in the atmosphere of those planets is properly modelled \citep{Hell14,Burn17}.  The C/H and C/O ratio should be good tracers of the formation mode of the planets and of their formation zone within the disks \citep[e.g.~][]{Hell09,Ober11,Ober16}. Comparing the formation zone to the present-day location of the planets could give clues on the efficiency of the migration processes and on past dynamical interactions, vital information needed to constrain and enrich planet formation theory \citep{Mord09}. 

The coupling of high contrast imaging and high-resolution spectroscopy offers additional prospects for detailed studies of the atmospheric physics of giant planets. Once the high resolution near-infrared spectrograph CRIRES at the VLT is back on sky in 2018 \citep[in~the~new~CRIRES+~mode~with~increased~sensitivity~and~wavelength~coverage,~][]{Dorn16}, it will be used as the backend for the SPHERE planet-imager, providing very high resolution star+exoplanet spectra.  This should enable the measurement of the rotation velocity of additional planets \citep{Sne14,Sne15} and will also provide an additional opportunity to apply retrieval methods to derive more accurate information on atmospheric states and chemical abundances. The high-resolution near-infrared spectrographs CARMENES \citep{Quir16} and SPIRou \citep{Arti14} will also provide spectra of tens of free-floating exoplanet analogues (young moving group members discussed previously). The high quality high resolution spectra of those objects will serve as proxies for understanding exoplanet physics. This may also allow us to determine if those objects were formerly companions that were ejected from nascent planetary systems due to dynamical interactions with other planets. 

\subsection{Prospects for detecting and characterizing habitable planets with direct imaging}

In the next few years, JWST will obtain transit spectroscopy for a handful of nearby M-star habitable zone planets potentially to be discovered by the TESS mission \citep{Rick10,Sull15}.  However, only a small percentage of habitable zone exoplanets are expected to transit.  In the next 10-20 years, direct imaging will eventually enable characterization of a larger sample of habitable zone exoplanets.  

\subsubsection{WISE 0855 -- the coolest atmosphere imaged to date}

\begin{figure}
\includegraphics[scale=.4]{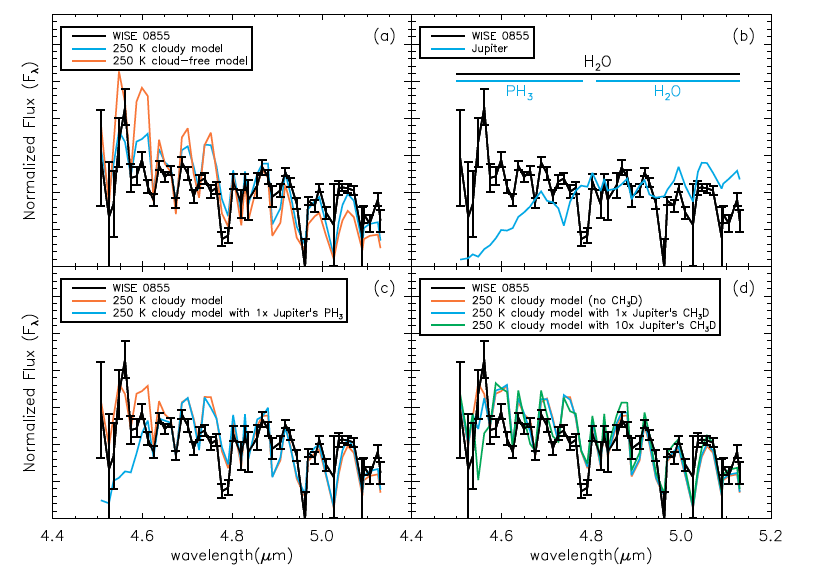}
\caption{Spectrum of the extremely cold ($T_\mathrm{eff} \sim 270 K$) brown dwarf WISE 0855, from \cite{Ske16}.  Water clouds are necessary to reproduce the observed spectral features.  The phosphine ($PH_3$) absorption features seen in Jupiter's spectrum are not apparent in WISE 0855.}
\label{fig:WISE0855}       
\end{figure}

While not likely habitable itself, the extremely cool, nearby brown dwarf WISE 0855 \citep{Luh14} has $T_\mathrm{eff}~\sim~270 K$, around the freezing point of water.  It is the coldest object ever imaged outside of our own solar system and gives us a sneak peak of what a habitable zone temperature atmosphere looks like.  \cite{Ske16} obtained the first spectrum of this object, at mid-IR wavelengths from 4.5 to 5.2 $\mu$m, presented here in Fig.~\ref{fig:WISE0855}.  The spectrum of this object shows clear evidence for the presence of water clouds, but does not show the strong $PH_3$ absorption seen in Jupiter's spectrum.  Given the presence of water clouds, \cite{Yat16} suggest that very cool brown dwarfs such as WISE 0855 may have an "atmospheric habitable zone" -- a range in its atmosphere with liquid water clouds as well as temperatures and pressures appropriate for Earth-centric life.

\subsubsection{Methods for Detecting and Characterizing Habitable Zone Planets}

Other than transit spectroscopy, three other methods will be able to characterize habitable zone exoplanets over the next decades: 

\begin{enumerate}
\item {\bf Thermal Phase Curves} -- Whether or not a planet transits, a tidally-locked planet may be detectable from its thermal phase curve -- the difference in temperature between the day side and night side of the planet, evident in the star-planet spectrum in the mid-IR \citep{Gaid04, Seag09, Sels11, Sels13, Maur12,Turb16,Krie16}.  
\item {\bf High-contrast imaging + high-resolution spectroscopy} The combination of high-resolution ground-based optical spectroscopy with high-contrast imaging can potentially characterize known planets \citep{Riau07, Sne15}.  First, high-contrast imaging is used to achieve $\sim 10^5$ contrasts in the inner arcsec, then high-resolution spectroscopy can distinguish between the stellar spectrum vs. the Doppler-shifted planet spectrum.  This technique has already been used to measure a rotation period for the young, giant exoplanet $\beta$ Pic b \citep{Sne14}.  \cite{Sne15} predict that a $R=1.5 R_{earth}$ planet with a twin-Earth thermal spectrum orbiting $\alpha$ Cen A could be detected in one night of E-ELT time using this method.
\item {\bf Direct imaging} Finally, in a decade, next-generation ground based extremely large telescopes (ELTs) as well as space coronagraphs such as WFIRST-AFTA will be able to detect very nearby habitable zone planets orbiting M stars \citep{Cro14b} as well as older gas giants around G stars (need citation).  In the longer term future, space missions such as LUVOIR will directly image habitable zone planets around G stars.
\end{enumerate}

\subsubsection{Proxima Centauri b}

2016 yielded the very exciting discovery of a habitable zone planet around Proxima Centauri \citep{Ang16}, the closest star to the Sun.  Proxima Centauri b has a minimum mass of 1.3 $M_{Earth}$ and an orbital period of 11 days, placing it squarely in the middle of the habitable zone around this low mass star.  While several groups have already modeled the potential habitability of this planet \citep{Riba16,Turb16,Barn16,Mead16,Gold16,Omal16}, as it has only been observed indirectly, whether or not the planet is habitable is an open question.  Additionally, there are a number of factors that make habitability less likely -- specifically, its somewhat higher mass than the Earth \citep[which~is~highly~dependent~on~orbital~inclination~with~possible~values~up~to~8~earth~masses,][]{Kane17}, the fact that it is likely tidally locked, as well as the much higher radiation environment expected for this planet around an active M star relative to our own fairly quiet G star \citep{Turb16, Garr16, Atri17,Dave16}.  (For an excellent recent review of work to date on potential habitability of M-star planets, see \cite{Shie16}.)  Thus, direct atmospheric characterization is vital to determine if the planet can have liquid water on its surface or perhaps even displays evidence for biosignatures.  The transit probability for Proxima Centauri b is only $1.5\%$ and initial optical searches have yielded no conclusive evidence of transits \citep{Kipp17}.  \cite{Krie16} model the possibility of detecting Proxima Cen b via its thermal light curves with JWST.  This is highly dependent on the atmosphere of the planet -- \cite{Krie16} predict that they can distinguish between a bare rocky planet and a planet with 35$\%$ heat redistribution to the nightside at the 4-$\sigma$ level, but very long integration times (on order of a few months) would be necessary to detect the 9.8 $\mu$m ozone band in the case of an Earth-like atmosphere.  Thus, direct imaging will likely provide the first opportunity to characterize this planet and determine if it is truly habitable.

The estimated star-planet contrast for Proxima Centauri b is $\sim 10^7$ in the thermal IR.  While this seems extreme, this is is still moderate compared to the $10^{10}$ contrast expected for a habitable zone planet observed in reflected light around a G star \cite{Tra07}.  The current generation of planet-imagers such as GPI and SPHERE already reach such contrasts -- but not at the separations necessary to image Proxima Centauri b, as Proxima Centauri b is expected to be within 40 mas of A \citep{Turb16}.  Thus, detecting Proxima Centauri b requires currently achievable contrasts, but diffraction-limited performance on a 30-40 m class telescope.  Likely in a decade, instruments such as METIS on the E-ELT will be the first to image Proxima Centauri b, if it is reflective enough.  However, while ELTs may detect Proxima Centauri b, they may not be suitable for the task of detecting a wide range of biosignatures and demonstrating dis-equilibrium chemistry.  As many potential biosignature gas features may also have abiotic sources \citep[e.g.~ozone~][]{Doma14,Luga15,Harm15}, coverage of multiple biosignatures and clear evidence of dis-equilibrium chemistry will be necessary to prove habitability.  Many such signatures are at UV wavelengths, inaccessible with ground-based ELTs. \cite{Luge16} suggest aurorae on Proxima Centauri b could also be detected via direct imaging with an extreme AO coronagraph on an ELT.  

In the shorter term, the technique of high-contrast imaging + high-resolution spectroscopy may be able to detect Proxima Centauri b, but only with considerable expenditure of observing time.  \cite{Lovi16} find 20-40 nights of telescope time with upgraded versions of the high-contrast imager SPHERE and the high-resolution spectrograph ESPRESSO would be required for a $5-\sigma$ detection of Proxima Centauri b, assuming an Earth-like atmosphere.  To search for biosignatures (the combination of $O_2$, water vapor, and methane) would require 60 nights of telescope time for a $3.6-\sigma$ detection. 

Proxima Centauri's higher mass neighbors may also be excellent targets to search for habitable zone planets.  \citet{Beli15} argue that because $\alpha$ Centauri A and B are so nearby and thus have a much higher apparent habitable zone size compared to any other FGKM stars in the sky, a habitable zone exoplanet around $\alpha$ Centauri A or B could be imaged with a 30-45 cm space telescope and advanced post-processing techniques.

\subsubsection{Other habitable zone exoplanets}

\begin{figure}
\includegraphics[scale=.5]{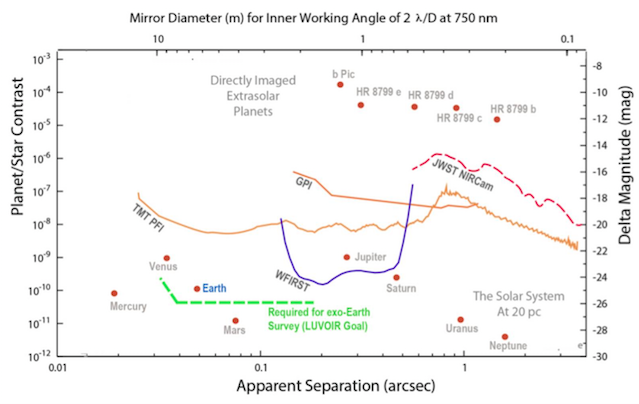}
\caption{Contrasts achieved by current exoplanet direct imaging instruments and predicted contrasts for next-generation images, from http://asd.gsfc.nasa.gov/luvoir, adapted from \citet{Laws12, Mawe12}.}
\label{fig:LUVOIR}       
\end{figure}

Most of the currently known habitable zone exoplanets \citep[largely~Kepler~discoveries~around~K~and~M~stars,][]{Kane16} will remain out of the reach of direct 
imaging, given their tight separations from their parent stars.  However, similar exoplanets around very nearby stars may be imaged with extremely large telescopes. Especially for planets around nearby M-stars, the limiting factor is not contrast, but resolution.  One way to probe even closer in to a star is to combine high-contrast imaging with high-resolution spectroscopy \citep{Sne15}.  As mentioned previously, an upgraded SPHERE + ESPRESSO at the VLT may be able to detect Proxima Centauri b in this manner \citep{Lovi16}.  Another way to increase resolution is simply to use a bigger telescope with adaptive optics correction to ensure diffraction-limited performance.  Thus, the advent of 30-40 m extremely large telescopes (ELTs) will pave the way for detection and characterization of giant exoplanets down to habitable zone exoplanets around M stars.  All three ELT projects eventually have plans for extreme-AO coronagraphs to detect planets \citep{Maci06,Kasp10,Davi16}.  \citet{Cro13} simulates the distribution of Kepler-discovered exoplanets around nearby, very low mass stars and then checked which planets were detectable using extremely large ground-based telescopes with high-contrast imagers and predicts that such a cohort may yield 2-4 directly-imageable planets with radii of 2-4 Earth radii and temperatures of $\sim$300 K -- in other words, super-Earths in the habitable zone.
\citet{Quan15} simulate performance for the E-ELT facility mid-IR instrument METIS and find that it will be able to image 1) $\sim$20 known RV Jovian exoplanets and 2) a handful of super-Earths in the habitable zone of nearby M-stars.  Combining high-contrast imaging with high-resolution spectroscopy with METIS, \citet{Sne15} predict that a $R=1.5 R_{earth}$ planet with a twin-Earth thermal spectrum orbiting $\alpha$ Cen A could be detected in one night of E-ELT time using this method, which also provides the resolution necessary to detect biosignature gas features in the spectrum of such an object.  Predicted contrasts for the planned TMT Planet Formation Imager (PFI) as well as other existing and future instruments are presented in Fig.~\ref{fig:LUVOIR}.

WFIRST-AFTA will be the first custom space coronagraph to directly image exoplanets \citep{Noec16}.  WFIRST-AFTA will directly image old RV planets down to super-Earths in the solar neighborhood, but, with a 2-m mirror diameter, will not have the contrast or resolution to image Earth twins (see Fig.~\ref{fig:LUVOIR} and also \citet{Robi16}).  \citet{Robi16} simulate expected WFIRST-AFTA coronagraph performance and conclude that WFIRST-AFTA will be able to detect methane absorption in cool Jupiters (Jupiters at 2 AU flux equivalent distance from the Sun) and obtain spectra of cool Neptunes as well as super-Earths.  Use of a star-shade with WFIRST-AFTA may eventually enable the detection and characterization of Earth-twins -- see e.g. the Exo-S mission concept \citep{Seag15}.

In the longer term future, planned missions such as HabEx and LUVOIR \citep{Dalc15} will combine large diameter space telescopes with state-of-the-art-coronagraphs, achieving the $10^{10}$ contrasts and extreme resolutions to detect and characterize low mass planets in the habitable zones around G stars.  In addition, LUVOIR is planned to have a wide wavelength coverage, from IR to UV.  As noted previously, as many single biosignature features may also have abiotic sources \citep[e.g.~ozone~][]{Doma14,Luga15,Harm15}, to truly determine if a planet is habitable requires the detection of a constellation of different biosignatures in the planet spectrum, the combination of which prove dis-equilibrium chemistry is present.  Additionally, to build a picture of habitability as a function of host star, planet mass, etc., it is vital to characterize habitable zone planets orbiting a wide range of host stars. In Figure~\ref{fig:LUVOIR}, contrasts achievable with current and future high resolution imagers are plotted alongside current known directly imaged exoplanets and our own solar system planets (assuming a distance of 20 pc). There are 64 G0-G5 stars within 20 pc of the Earth. While earlier missions and telescopes such as JWST, WFIRST, and the ELTs will yield photometry and spectroscopy of a handful of habitable zone planets around low mass M and K dwarfs (e.g. Proxima Centauri b), only a mission like LUVOIR will yield the contrasts and resolutions necessary to detect and characterize true Earth twins (i.e. Earth-sized planets in the habitable zone of Sun-like stars). Through a LUVOIR-class mission we will be able to observe enough planets to reach statistically meaningful conclusions regarding the frequency and characteristics of exo-Earths and better understand our own planet in its astronomical context.

\section{Conclusions}

Direct imaging has already transformed our understanding of young giant planets, both as companions and free-floating in young moving groups.  Such low-gravity objects have much different atmospheric structures than high-gravity, field brown dwarfs with similar $T_\mathrm{eff}$, holding on to thick dusty clouds in their atmospheres and suppressing methane absorption down to much lower temperatures than their higher mass counterparts.  Ultimately, current work in direct imaging is a prelude to the era of deep characterization that will begin with JWST and find its ultimate expression in future space telescopes such as LUVOIR.  JWST will enable in-depth characterization of young giant exoplanets, while the ELTs will allow for the discovery and characterization of old, cold Jupiters as well as habitable zone exoplanets around very nearby M-stars.  Eventually, a LUVOIR-class mission will be able to obtain images and spectroscopy of enough habitable zone exo-Earths to reach statistically meaningful conclusions regarding the frequency and characteristics of these planets.  Perhaps such a mission will conclusively demonstrate the presence of dis-equilibrium chemistry and show evidence of not just being in the habitable zone but actually being inhabited.

\section{Cross-References}

\begin{itemize}
\item{Variability of Brown Dwarfs}
\item{Large Scale Searchers for Brown Dwarfs and Free-Floating Planets}
\item{Direct Imaging as an Exoplanet Discovery Method}
\item{Imaging with Adaptive Optics and Coronographs for Exoplanet Research}
\item{SPHERE as an Instrument for Exoplanet Research}
\item{CHARIS on Subaru: An Instrument for Exoplanet Research}
\item{ESPRESSO on VLT: An Instrument for Exoplanet Research}
\item{Space Missions for Exoplanet Science: JWST}
\item{Atmospheric Retrieval for Exoplanet Atmospheres}
\item{Atmospheres as the Window Into Exoplanet Habitability and Life}
\item{HR8799: Imaging a System of Exoplanets}
\item{Future Exoplanet Research: High Contrast Imaging Techniques}
\end{itemize}

\begin{acknowledgement}
BB gratefully acknowledges support from STFC grant ST/M001229/1.
\end{acknowledgement}

\bibliographystyle{spbasicHBexo}  
\bibliography{HBexoTemplateBib} 

\end{document}